\documentclass[a4paper,11pt]{article}
\usepackage{pos}

\usepackage{booktabs,array}
\usepackage[utf8]{inputenc}
\usepackage{float,graphicx,hhline}
\usepackage{algorithm,algpseudocode}
\usepackage{appendix}
\usepackage{mathtools}
\usepackage{multirow}
\usepackage{slashed}
\usepackage{pifont}
\usepackage[caption=false]{subfig}
\usepackage[export]{adjustbox}
\usepackage{enumitem} 
\usepackage{bbm}
\usepackage{dsfont}
\usepackage{hyphenat}
\usepackage{braket}
\usepackage{comment}

\usepackage{hyperref}
\hypersetup{
    pdftitle={},
    colorlinks=true,     
    linkcolor=blue,      
    citecolor=blue,      
    filecolor=blue,      
    urlcolor=blue        
}
\usepackage[capitalize]{cleveref}
\crefname{equation}{Eq.}{Eqs.}
\crefname{section}{Sec.}{Secs.}

\DeclareMathOperator{\Tr}{Tr}

\title{Practical applications of machine-learned flows on gauge fields}

\author[b,c]{Ryan~Abbott}
\author[d]{Michael~S.~Albergo}
\author[b,c]{Denis~Boyda}
\author*[a,b,c]{Daniel~C.~Hackett}
\author[e]{Gurtej~Kanwar}
\author[b,c]{Fernando~Romero-L\'opez}
\author[b,c]{Phiala~E.~Shanahan}
\author[b,c]{Julian~M.~Urban}

\affiliation[a]{Fermi National Accelerator Laboratory, Batavia, IL 60510, U.S.A.}
\affiliation[b]{Center for Theoretical Physics, Massachusetts Institute of Technology, Cambridge, MA 02139, USA}
\affiliation[c]{The NSF AI Institute for Artificial Intelligence and Fundamental Interactions}
\affiliation[d]{Center for Cosmology and Particle Physics, New York University, New York, NY 10003, USA}
\affiliation[e]{Albert Einstein Center, Institute for Theoretical Physics, University of Bern, 3012 Bern, Switzerland}

\emailAdd{dhackett@fnal.gov}

\abstract{
Normalizing flows are machine-learned maps between different lattice theories which can be used as components in exact sampling and inference schemes. Ongoing work yields increasingly expressive flows on gauge fields, but it remains an open question how flows can improve lattice QCD at state-of-the-art scales. We discuss and demonstrate two applications of flows in replica exchange (parallel tempering) sampling, aimed at improving topological mixing, which are viable with iterative improvements upon presently available flows.
}

\FullConference{The 40th International Symposium on Lattice Field Theory (Lattice 2023)\\
July 31st - August 4th, 2023\\
Fermi National Accelerator Laboratory\\}

\begin{document}

\maketitle

\section{Introduction}

Numerical lattice quantum chromodynamics (QCD) is an integral part of the modern particle and nuclear theory toolkit~\cite{morningstar2007,USQCD:2019hyg,Lehner:2019wvv,Kronfeld:2019nfb,Cirigliano:2019jig,Detmold:2019ghl,Bazavov:2019lgz,Joo:2019byq,Boyle:2022uba}.
In this framework, the discretized path integral is computed  using Monte Carlo methods.
Computationally, this is very expensive, and grows more so as physical limits of interest are approached~\cite{Wolff:1989wq,Schaefer:2009xx,Schaefer:2010hu}.
Consequently, algorithmic developments are an important driver of progress.
For example, resolving topological freezing~\cite{Alles:1996vn,DelDebbio:2004xh,Schaefer:2010hu}---an issue that arises in sampling gauge field configurations with state-of-the-art Markov chain Monte Carlo (MCMC) algorithms like heatbath~\cite{Creutz:1980zw, Cabibbo:1982zn, Kennedy:1985nu, Brown:1987rra, Adler:1987ce} or Hybrid/Hamiltonian Monte Carlo (HMC)~\cite{Duane:1987de,neal1993probabilistic,neal1996bayesian}---would provide access to finer lattice spacings than presently affordable.
To such ends, recent work has explored how emerging machine learning (ML) techniques may be applied to lattice QCD~\cite{Boyda:2022nmh, Cranmer:2023xbe}.
Of particular interest has been the possibility of accelerating gauge-field sampling~\cite{Kanwar:2020xzo,Boyda:2020hsi,Foreman:2021ixr,Foreman:2021ljl,Foreman:2021rhs,Albergo:2022qfi,Finkenrath:2022ogg,Abbott:2022zhs,Abbott:2022hkm,Abbott:2023thq} using normalizing flows~\cite{rezende2016variational,dinh2017density,JMLR:v22:19-1028}, a class of generative statistical models with tractable density functions.

In this framework, a flow $f$ is a learned, invertible (diffeomorphic) map between gauge fields.
Abstractly, flows may be thought of as bridges between different distributions over gauge fields (or, equivalently, different theories or choices of action parameters).
When applied to samples $V$ from some chosen base distribution $r(V)$ as $U = f(V)$, the flow induces a model distribution $q(U)$, whose density may be computed exactly as
\begin{equation}
    q(U) = r(V) ~ / ~ J_f(V) ~ , 
\end{equation}
where $J_f(V) = | \det \partial f(V) / \partial V |$ is the Jacobian determinant of the flow transformation.
The flow is an \emph{exact} bridge between the base and model distributions: ensembles distributed per $r$ may be transformed to ensembles distributed per $q$ and vice versa.
However, the form of the model distribution is a complicated function of the architecture and precise parameter values of the learned flow and does not typically correspond to a distribution of physical interest.

Instead, for sampling applications, one typically variationally optimizes the flow so that $q$ approximates some target density of interest $p$.
In this case, the flow serves as an \emph{approximate} bridge between the base $r$ and target $p$.
Because the density of $q$ can be evaluated, this approximation can be corrected to obtain exact results for $p$.
Samples drawn from $r$ and flowed are samples from $q$, which can
be used to compute reweighted expectations under the target as $\braket{O}_p = \braket{pO/q}_q$
or used as proposals in independence Metropolis to construct a $p$-distributed Markov chain~\cite{Albergo:2019eim,noe2019boltzmann,Metropolis:1953am,Hastings:1970aa,tierney1994markov}.

Most demonstrations thus far have been in the context of
the simplest application of such an approximate bridge: full generative modeling or ``(approximate) direct sampling''.
In this case, the flow bridges between a Haar uniform base $r(U)=1$ and the target theory of interest $p$, specified in terms of an action as $p(U) \propto \exp[-S_p(U)]$.
If achieved for QCD at scale, this  would offer significant advantages over the present state of the art, including embarrassingly parallel sampling and a natural resolution of topological freezing~\cite{Cranmer:2023xbe}.
However, this requires higher-quality models than presently available.

This raises an important question: how can presently available flows be used to improve lattice QCD?
The key idea is that we are free to use a flow to bridge between any $r$ and $p$ of our choosing.
Neither distribution need be trivial, and choosing more similar distributions makes for a substantially easier modeling task.
In this Proceedings, we explore two promising applications following this idea.
After detailing our numerical setup in \cref{sec:setup}, we provide preliminary numerical demonstrations in \cref{sec:t-rex,sec:dr-rex}
of how flows may accelerate sampling of QCD field configurations using the parallel tempering / replica exchange algorithm.
The first in \cref{sec:t-rex} uses flows to bridge nearby physical theories to accelerate sampling in all of them.
The second in \cref{sec:dr-rex} uses flows to repair unphysical action defects introduced to accelerate topological mixing.
A recent publication, Ref.~\cite{fh}, presents a related but distinct application: flows can be employed to generate correlated ensembles for multiple different target theories, enabling correlated noise cancellation in computations of derivatives with respect to lattice action parameters.\footnote{Preliminary results for this strategy were presented in the talk corresponding to this Proceedings.}

\section{Numerical setup}
\label{sec:setup}

For the two numerical demonstrations below we consider pure-gauge $\mathrm{SU}(3)$ defined with the Wilson gauge action~\cite{Wilson:1974sk}, 
\begin{equation}
    S_g(U) = -\frac{\beta}{N_c} \sum_x \sum_{\mu<\nu} \mathrm{Re} \Tr P_{\mu\nu}(x) ~ ,
\end{equation} 
where $P_{\mu\nu}(x)$ is the plaquette with extent in the $\mu$ and $\nu$ directions.
We emphasize that flowed approaches to fermions and the fermion determinant have already been explored~\cite{Albergo:2021bna,Albergo:2022qfi,Abbott:2022zhs}, including in QCD~\cite{Abbott:2022hkm}.
These methods may be applied to generalize everything presented here to QCD. 
The choice to test in pure gauge theory is made only for simplicity and to reduce experimentation costs.

In both demonstrations, the objective is to use flows to improve topological mixing in sampling of pure-gauge targets.
The baseline to be improved upon is alternating hits of (pseudo)heatbath (HB) and (pseudo)overrelaxation (OR), denoted  as HB+OR~\cite{Creutz:1980zw, Cabibbo:1982zn, Kennedy:1985nu, Brown:1987rra, Adler:1987ce}.
We compute topological charges using the clover definition at Wilson flow time $t/a^2 = 2$~\cite{Luscher:2009eq,Luscher:2010iy,Narayanan:2006rf,Lohmayer:2012hs}.

MCMC sampling efficiency is typically assessed using the integrated autocorrelation time.
For some observable $O$, it is defined in terms of the autocorrelation function,
$C_O(\delta t) = 
 \braket{O(\delta t) O(0)} - \braket{O(\delta t)} \braket{O(0)}$,
where equilibration is assumed and the expectations are over all possible Markov chains, as 
\begin{equation}
    \tau_\mathrm{int}(O) = \frac{1}{2} + \sum_{\delta t=1}^{\infty} C_O(\delta t) ~ .
\end{equation}
In practice, the sum must be truncated at some $t_\mathrm{max} < \infty$ due to noise in $C_O$ estimates at large $\delta t$.
Autocorrelations increase the variance of estimators of $O$ by a factor of $2 \tau_\mathrm{int}(O)$ versus independent samples with the minimal $\tau_\mathrm{int} = 1/2$~\cite{Wolff:2003sm}.
Thus, computational cost is linear in $\tau_\mathrm{int}$.

The model architectures used in both demonstrations are variations on the residual flows~\cite{Abbott:2023thq} used in Ref.~\cite{fh}.
This ``coupling layer'' architecture uses variable partitioning to guarantee invertibility and a tractable Jacobian~\cite{dinh2014nice,dinh2017density}.
Schematically, each layer updates a subset of ``active links'' conditioned on the other ``frozen'' ones by applying a step of gradient flow with respect to a learned potential (cf.~the Wilson gauge action in Wilson flow).
In each layer, active links are chosen as those in one direction and for one checkerboard parity over sites.
The models here have 16 layers such that each link in the lattice is updated exactly twice.
Our learned potential is parametrized by applying a learned equivariant smearing to the frozen link field, then summing over plaquettes each constructed out of one active link and a frozen smeared staple as in Ref.~\cite[Sec.~2C]{fh}.
Certain modifications are required for the defect repair architecture in \cref{sec:dr-rex}, as discussed therein.

We use reverse KL self-training~\cite{Li2018,Albergo:2019eim,Kullback:1951} with path gradients~\cite{Abbott:2023thq,vaitl2022gradients} to optimize the models.
Note that the base distributions here are pure gauge theories at $\beta > 0$, rather than Haar uniform ($\beta=0$).
This does not modify the training procedure formally, but does require sampling the base distribution using HB+OR.
In practice, we accomplish this by evolving $B$ independent Markov chains alongside training, where $B$ is the batch size.
In some cases we also employed transfer learning between different volumes and action parameters~\cite{Abbott:2022zsh}. Further details of the ML methodology are provided in Ref.~\cite{fh}.

\section{Transformed Replica EXchange (T-REX)}
\label{sec:t-rex}

The parallel tempering or Replica EXchange (REX) algorithm~\cite{Swendsen:1986vqb,earl2005parallel} simultaneously samples parallel Markov chains for a sequence of different, overlapping target densities $(p_0, p_1, \ldots)$ on the same lattice geometry.
The algorithm proceeds by evolving the chains independently with an appropriate MCMC update for each target (here, HB+OR for the $\beta$ of that chain), and then occasionally proposing swaps of configurations between neighboring chains.
Intuitively, this swapping allows mixing with faster-moving targets to accelerate evolution in slower ones.
However, the swap acceptance rate (AR) falls as the lattice size grows, requiring increasingly many chains interpolating between any two target $\beta$s to maintain a finite swap AR.

As introduced in Ref.~\cite{invernizzi2022skipping}, a flow which approximately bridges two neighboring densities can be used to improve the swap AR between them, allowing more widely separated targets.
For a flow $f$, a proposed swap of $(U_0,U_1)$ between the chains for targets $p_0$ and $p_1$ proceeds as: flow each sample ``across the bridge'' as $(U_1', U_0') = (f^{-1}(U_1), f(U_0))$, then accept or reject the swap with probability\footnote{This is equivalent to the expression in Ref.~\cite{invernizzi2022skipping}.}
\begin{equation}
    p_\mathrm{acc} = \min\left[1,
        \frac{p_0(U_1') p_1(U_0')}{p_0(U_0) p_1(U_1)} J_f(U_0) J_{f^{-1}}(U_1)
    \right] ~ .
\end{equation}
If accepted, the next state is $(U_1', U_0')$; if rejected, the present state $(U_0,U_1)$ is replicated.
We call this algorithm ``Transformed Replica EXchange'' or T-REX.
Several limiting cases are worth noting.
For an exact flow between $p_0$ and $p_1$, $p_\mathrm{acc} = 1$.
For an identity flow, $f(U) = U$, untransformed/standard REX is recovered.
For two chains, one targeting Haar uniform and the other the theory of interest, and with no MCMC update for the target chain,\footnote{This amounts to the choice of the identity operation as the MCMC update for the target stream. Using a nontrivial update for the target stream is equivalent to alternating MCMC updates with flow model proposals as explored in Ref.~\cite{Hackett:2021idh}.}
we recover a sampling scheme equivalent to approximate direct sampling with independence Metropolis.

\begin{figure}
    \centering
    \includegraphics[width=\linewidth]{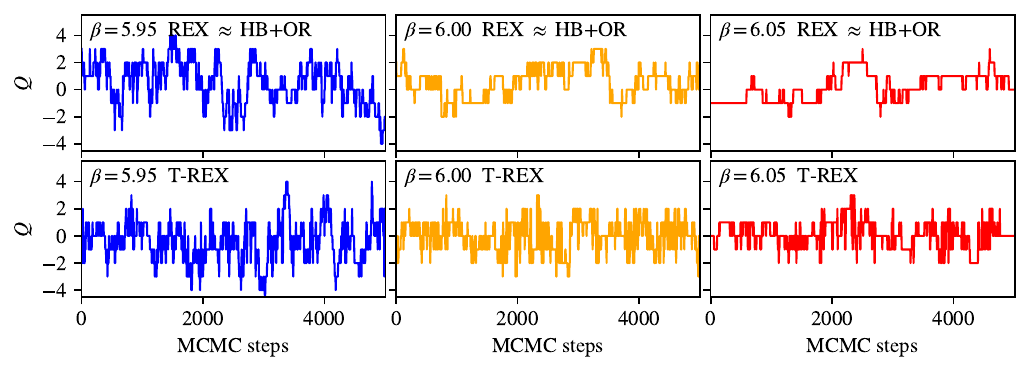}
    \caption{
        Evolution of the topological charge for a set of three different pure-gauge targets on a $12^4$ volume, sampled simultaneously using either REX or T-REX.
        REX swaps are almost never accepted, so it is  equivalent to independent HB+OR on each stream.
        One MCMC step is 5 HB hits followed by 2 OR hits.
        Swaps are proposed every 5 steps.
    }
    \label{fig:rex-vs-trex-Q}
\end{figure}

\begin{figure}
    \centering
    \includegraphics[width=\linewidth]{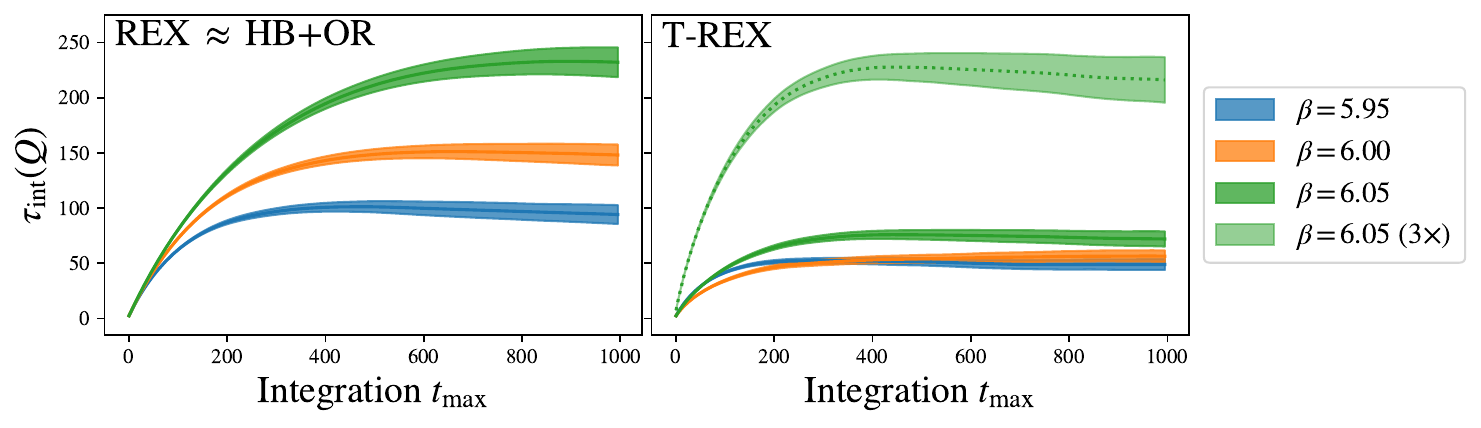}
    \caption{
        Running estimates of the integrated autocorrelation times of the topological charge for three pure-gauge actions, sampled simultaneously using either REX or T-REX as in \cref{fig:rex-vs-trex-Q}.
        Units are MCMC steps as in \cref{fig:rex-vs-trex-Q}.
        The curves show $1/2 + \sum_{\delta_t=1}^{t_\mathrm{max}} C_Q(\delta t)$ and are expected to plateau at $\tau_\mathrm{int}(Q)$.
        Uncertainties are evaluated as the standard error over 32 (40) independent streams for REX (T-REX), each of length 10000 (4500) steps.
        For T-REX, the $\beta=6.05$ curve is reproduced with a factor of 3 for ease of cost comparison.
    }
    \label{fig:rex-vs-trex-ac}
\end{figure}

For a numerical test, we consider simultaneously sampling three chains with $\beta = (5.95, 6, 6.05)$ on a $12^4$ lattice geometry.
Applying untransformed REX, we find the swap AR is  no more than $10^{-4}$ for both $5.95 \leftrightarrow 6$ and $6 \leftrightarrow 6.05$.
To improve this, we train two independent flows using the setup of \cref{sec:setup}: one to bridge $5.95 \leftrightarrow 6$ and another for $6 \leftrightarrow 6.05$.
Note that we train on $4^4$ lattices before transferring to $12^4$ lattices, and also transfer between target parameters rather than training each flow from scratch.
Applying T-REX with these flows, we obtain swap ARs of $\approx 15\%$ and $20\%$, respectively.

\cref{fig:rex-vs-trex-Q} compares MCMC histories of the topological charge on the stream for each $\beta$, as sampled using untransformed REX $\approx$ HB+OR (equivalent here due to the $\approx$ zero swap AR) and T-REX using the aforementioned flows.
The onset of topological freezing is clearly visible in the untransformed REX streams.
T-REX provides a clear improvement in the mixing rate in all three streams.

To be more quantitative, in \cref{fig:rex-vs-trex-ac} we estimate the integrated autocorrelation time $\tau_\mathrm{AC}$ of the topological charge in each stream.
Even the slowest-moving chain ($\beta=6.05$) in T-REX is faster than the fastest chain ($\beta=5.95$) with REX $\approx$ HB+OR.
Whether this represents an advantage depends on the computational goal.
We consider two scenarios.
First, in a multi-ensemble calculation this is an unambiguous speed-up if flow costs can be neglected: 
for the same number of HB+OR steps, there is a gain in effective statistics in every stream.
In addition, due to the swapping, the three T-REX ensembles are correlated, which may be useful as discussed in Ref.~\cite{fh}.
Second, one might imagine only $\beta=6.05$ is of interest, and the goal is to resolve topological freezing by bridging to lower, unfrozen $\beta$s as in approximate direct sampling.
In this case, the other $\beta$s are sampled only to improve mixing and discarded after as a side-product.
We find that after paying the $3\times$ overhead factor of running HB+OR for three streams instead of one (and again neglecting flow costs), the cost of sampling $\beta=6.05$ is the same with either HB+OR or T-REX.
Apparently, it will be necessary to bridge more widely separated $\beta$s to obtain an advantage in this scenario.

\section{Defect Repair Replica EXchange (DR-REX)}
\label{sec:dr-rex}

The different targets in replica exchange need not all be theories of physical interest.
This allows for opportunistic choices of unphysical actions designed to accelerate topological charge mixing when sampled alongside a single physical target stream.
An existing approach which exploits this is Parallel Tempering Boundary Conditions (PTBC)~\cite{Hasenbusch:2017unr,Bonanno:2020hht,Bonanno:2022yjr}.
The idea is as follows.
Open boundary conditions (OBCs) are known to alleviate topological freezing~\cite{Luscher:2011kk}, but periodic boundary conditions (PBCs) are preferable otherwise.
PTBC uses REX to sample a sequence of actions interpolating between OBCs and PBCs to accelerate topological mixing in the target PBC stream.
In practice, rather than using OBCs across the full spatial extent of the lattice, introducing a localized OBC defect (``poking a hole in the boundary'') can be sufficient to speed up topological mixing while requiring fewer interpolating chains.

More precisely, these defected theories are defined here using a generalization of the Wilson pure gauge action with a separate $\beta$ for each plaquette, i.e.,
\begin{equation}
    S_g(U) = -\frac{1}{N_c} \sum_x \sum_{\mu<\nu} \beta_{\mu\nu}(x) \, \mathrm{Re} \Tr P_{\mu\nu}(x) ~.
\end{equation}
An OBC-type defect is defined as one where $\beta_{\mu\nu}(x) = \beta_d < \beta$ for all defect plaquettes, as defined by those touching the timelike links in a $L_d^3 \times 1$ volume, and $\beta_{\mu\nu}(x) = \beta$ for all other plaquettes. True temporal OBCs are recovered for a full-lattice defect $L_d = L$ and $\beta_d=0$.
Other defect geometries can be considered and may work better in practice.
HB+OR can be generalized straightforwardly to sample these actions.

The novel approach explored here is to apply T-REX to PTBC.
In this case, the role of the flow is to (perhaps partially) repair the defect, hence we dub the algorithm Defect Repair Replica EXchange (DR-REX).
The locality of the defect makes this a particularly appealing application.
Because the physical effects of the defect are localized in its neighborhood, the defect repair flow needs only to act on a subvolume containing this neighborhood, formally performing the identity operation elsewhere.
This confers two advantages.
First, the swap AR given some subvolume flow is approximately the same for any target volume, with no exponential scaling~\cite{Finkenrath:2022ogg,Abbott:2022zsh}.
Second, the computational cost of applying the subvolume flow is independent of the target volume, such that applying the flow can be much less expensive than the MCMC updates for large-volume targets.

\begin{figure}
    \centering
    \hfill
    \subfloat{{
    \includegraphics[height=1.5in]{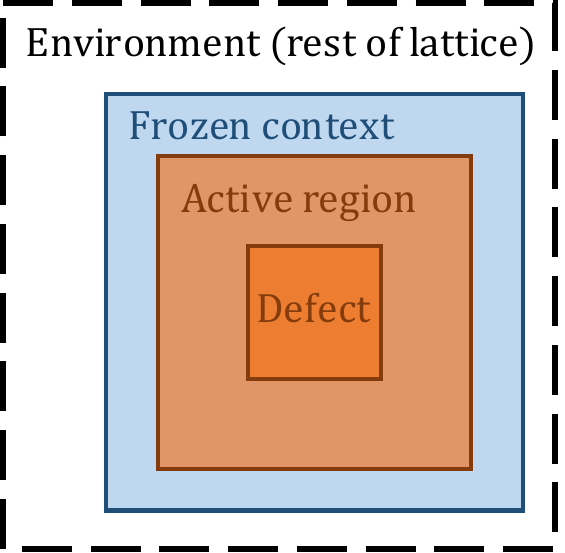}
    }}
    \! \hfill \!
    \subfloat{{
    \includegraphics[height=1.5in]{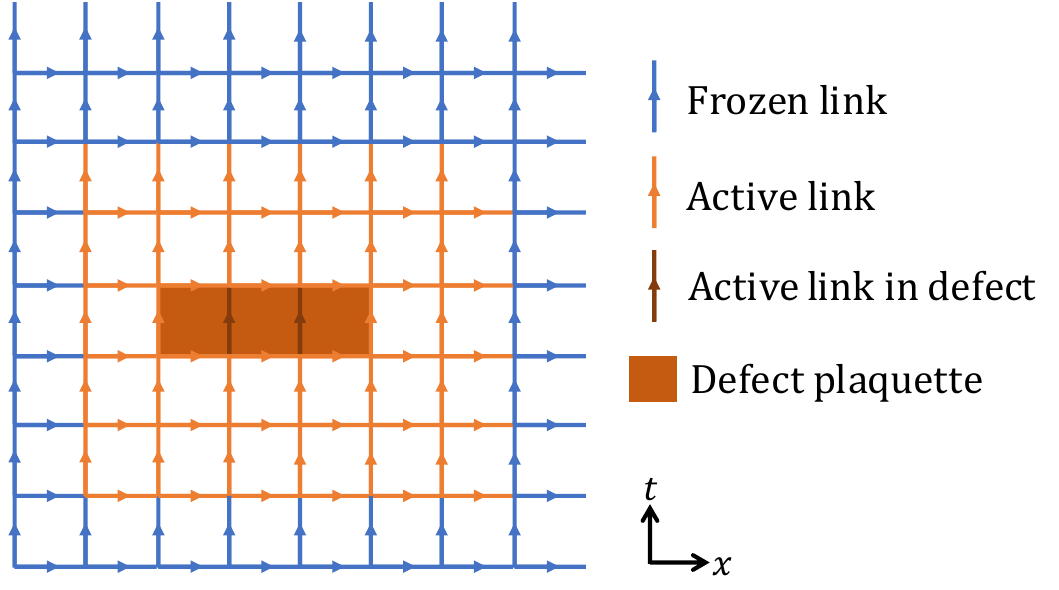}
    }}
    \hfill
    \!
    \caption{Sketches of the geometry for defect repair flows.
    At left, the flow performs the identity operation outside the orange region. The active region around the defect (orange) is acted upon, conditioned on its local context (blue).
    At right, the exact subvolume geometry used for the flows in this work. 
    This geometry is symmetric in all spatial directions ($y$, $z$ not shown).
    }
    \label{fig:defect-sketch}
\end{figure}

The geometry acted on by the flow is sketched in \cref{fig:defect-sketch}.
A halo of links at the boundary of the subvolume is kept frozen throughout the flow.
The absence of translational symmetry in the problem requires generalizing the architecture used in the previous application.
We use the straightforward option of promoting every parameter in the model to a lattice of parameters, i.e.~adding a site index to each parameter.\footnote{This is analogous to generalizing convolutional layers to define ``locally connected layers''.}
We break reflection symmetries in the model in a similar manner.
Although this massively increases the parameter count, the computational cost of training or applying the flow is not substantially increased.
Note that while this choice ``hard codes'' a flow to a specific subvolume geometry, it can still be applied to subvolumes embedded in different target volumes.
Subvolume-transferrable models could be obtained by e.g.~instead parametrizing this site dependence with neural networks, but we leave these complications for future work.

\begin{figure}
    \centering
    \includegraphics[width=\linewidth]{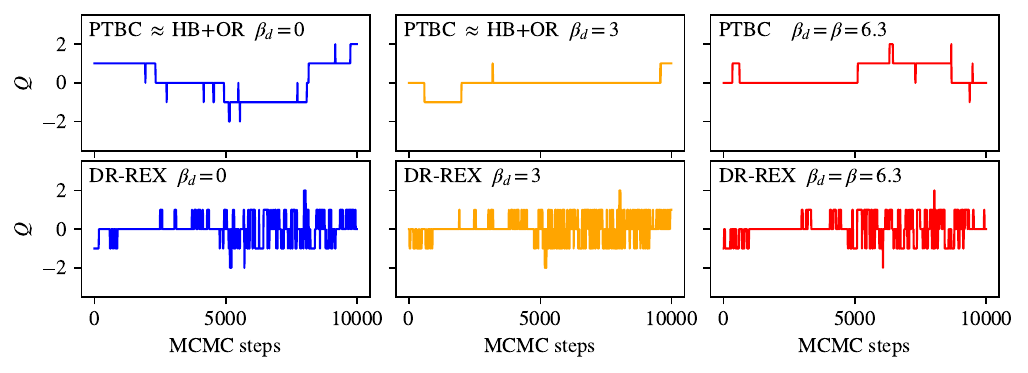}
    \caption{
        Evolution of the topological charge under PTBC or DR-REX for pure-gauge targets with $\beta=6.3$ on a $16^4$ volume with OBC defects of size $2^3$ with $\beta_d=0,3,6.3$, where $\beta_d=\beta=6.3$ is the target stream with no defect.
        PTBC swaps are almost never accepted, so it is equivalent to independent HB+OR on each stream.
        One MCMC step is 1 HB hit followed by 5 OR hits.
        Swaps are proposed every 10 steps.
    }
    \label{fig:ptbc-vs-drrex-Q}
\end{figure}

\begin{figure}
    \centering
    \includegraphics[width=\linewidth]{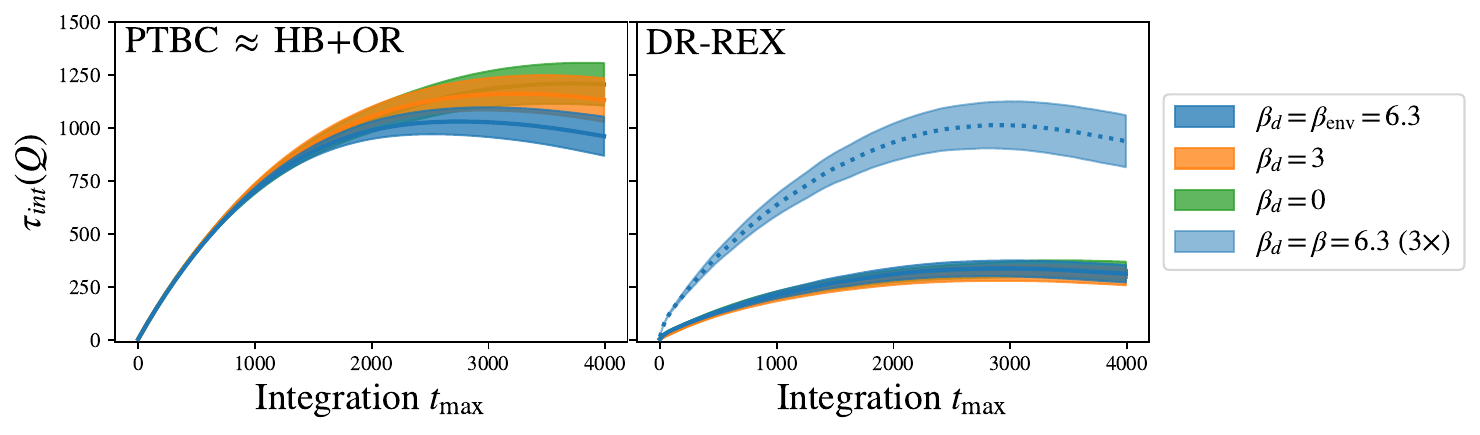}
    \caption{
        Running estimates of the integrated autocorrelation times of the topological charge for sampling using PTBC or DR-REX as in \cref{fig:ptbc-vs-drrex-Q}.
        Units are MCMC steps as in \cref{fig:ptbc-vs-drrex-Q}.
        The curves show $1/2 + \sum_{\delta_t=1}^{t_\mathrm{max}} C_Q(\delta t)$ and are expected to plateau at $\tau_\mathrm{int}(Q)$.
        Uncertainties are evaluated as the standard error over 24 independent streams, each of length 29000 steps.
        For DR-REX, the $\beta_d=6.3$ curve is reproduced with a factor of 3 for ease of cost comparison.
    }
    \label{fig:ptbc-vs-drrex-ac}
\end{figure}

For our numerical test, we consider OBC defects of size $2^3$ in a $16^4$ target at $\beta=6.3$.
We train two defect repair flows which act on $8^4$ subvolumes, one bridging between $\beta_d=0$ and 3, and a second bridging 3 and 6.3, where $\beta_d=6.3$ is the undefected target.
\Cref{fig:defect-sketch} shows the precise subvolume geometry.
We train each flow independently, using smaller volumes ($10^4$, then $12^4$) than the target.
Sampling these three theories with $\beta_d=0,3,6.3$ with untransformed PTBC results in a very small swap rate; we observed no accepted swaps in $10^5$ attempts.
Adding the flows enables swap ARs of 23\% and 28\% in DR-REX, thus providing at least an $O(10^4)$ improvement.
A more fair comparison may be that unflowed PTBC obtains comparable swap ARs with $\beta_d$ interpolated over 7-8 chains, amounting to a $\sim 2\times$ improvement.

\Cref{fig:ptbc-vs-drrex-Q} compares topological charge histories for PTBC $\approx$ HB+OR with those of DR-REX.
Clearly, the topological charge is severely frozen for HB+OR, and changes more rapidly under DR-REX.
However, much of this apparent mixing is due only to swapping, rather than tunneling events.
To assess whether an advantage has been obtained, we estimate $\tau_Q^\mathrm{int}$ in \cref{fig:ptbc-vs-drrex-ac}.
Accounting for the $3 \times$ overhead of running HB+OR for three chains, we find that DR-REX is only at the break-even point with HB+OR, neglecting flow costs.
Evidently, the $2^3$ OBC defect does not sufficiently increase the tunneling rate (note $\tau_Q^\mathrm{int}$ is approximately the same across the different HB+OR streams).
We conclude that it will be necessary to repair larger defects to obtain a true computational advantage.

\section{Outlook}
\label{sec:conclusion}

There are a number of promising avenues to employ machine-learned flows on gauge fields for computational advantages in lattice QCD calculations.
Ref.~\cite{fh} demonstrated uses of correlated ensembles generated by flows.
In this Proceedings, we explore two different replica exchange methods, T-REX and DR-REX, to improve topological freezing.
Computational advantage has yet to be shown when accounting for all costs associated with the learned components. However, if they may be neglected, there are already demonstrated advantages if several nearby ensembles are to be generated simultaneously. Moreover, there are structural reasons to believe these methods are especially promising.

Flowed replica exchange methods may be thought of as a generalization of approximate direct sampling with an adjustable trade-off between how much of the sampling task is accomplished by flows versus traditional MCMC updates.
With two chains, T-REX provides an exact sampling scheme to use flows between a target theory and a nontrivial base distribution, even when exact sampling is not possible for the base distribution.
This can enable useful flow-based sampling before full trivializing maps are available, and a smooth approach to the capabilities of approximate direct sampling as flow technology develops further.
Using more chains to interpolate over a sequence of theories further eases requirements on flow model quality.
We note that these same advantages also hold for the closely related class of algorithms including Continual Repeated Annealed Flow Transport Monte Carlo (CRAFT)~\cite{Matthews:2022sds,arbel2021annealed} and various limiting cases thereof~\cite{neal2001annealed}, including stochastic normalizing flows (SNFs)~\cite{wu2020stochastic,Caselle:2022acb}.

The combination of flows with the PTBC algorithm, DR-REX, has important structural advantages that make it an especially promising application of flows.
Flows can increase the swap AR between chains, potentially making larger defects practical than with untransformed replica exchange alone.
Furthermore, defect repair flows which act on a fixed-size embedded subvolume can be applied to larger target volumes without increasing the application cost of the flow or the required model quality.
The defects treated in the application here were too small to alleviate freezing, via either PTBC or DR-REX, but there is a clear path forward with well-defined learning tasks achievable with iterative improvements over what is done here.

With further model tuning (which was not extensively explored here), and continued development of flow technology, both T-REX and DR-REX present promising avenues to computational advantage in lattice QCD gauge field generation at scale. To fully exploit novel flow technology it will be important to further develop these methods and continue to explore other algorithmic approaches to accelerate sampling by incorporating flows.

\section*{Acknowledgements}

We thank Aleksandar Botev, Kyle Cranmer, Alexander G.~D.~G.\ Matthews, S\'ebastien Racani\`ere, Ali Razavi, and Danilo J. Rezende for useful discussions and valuable contributions to the early stages of this work. RA, DCH, FRL, PES, and JMU are supported in part by the U.S.\ Department of Energy, Office of Science, Office of Nuclear Physics, under grant Contract Number DE-SC0011090. PES is additionally supported by the U.S.\ DOE Early Career Award DE-SC0021006, by a NEC research award, and by the Carl G and Shirley Sontheimer Research Fund. FRL acknowledges support by the Mauricio and Carlota Botton Fellowship. GK was supported by the Swiss National Science Foundation (SNSF) under grant 200020\_200424. This manuscript has been authored by Fermi Research Alliance, LLC under Contract No.~DE-AC02-07CH11359 with the U.S. Department of Energy, Office of Science, Office of High Energy Physics. MSA is supported by the National Science Foundation under the award PHY-2141336. MSA thanks the Flatiron Institute for their hospitality. This work is supported by the U.S.\ National Science Foundation under Cooperative Agreement PHY-2019786 (The NSF AI Institute for Artificial Intelligence and Fundamental Interactions, \url{http://iaifi.org/}) and is associated with an ALCF Aurora Early Science Program project, and used resources of the Argonne Leadership Computing Facility which is a DOE Office of Science User Facility supported under Contract DEAC02-06CH11357. The authors acknowledge the MIT SuperCloud and Lincoln Laboratory Supercomputing Center~\cite{reuther2018interactive} for providing HPC resources that have contributed to the research results reported within this paper. Numerical experiments and data analysis used PyTorch~\cite{NEURIPS2019_9015}, JAX~\cite{jax2018github}, Haiku~\cite{haiku2020github}, Horovod~\cite{sergeev2018horovod}, NumPy~\cite{harris2020array}, and SciPy~\cite{2020SciPy-NMeth}. Figures were produced using matplotlib~\cite{Hunter:2007}.

\bibliographystyle{JHEP}
\bibliography{main}

\end{document}